\documentclass[aps,twocolumn,superscriptaddress]{revtex4-1}
\usepackage[colorlinks=true,bookmarks=true,citecolor=magenta,urlcolor=magenta,linkcolor=magenta,breaklinks]{hyperref} % command used
\usepackage{breakurl}
\usepackage{sidecap}
\usepackage{amssymb}
\usepackage{hhline}
\usepackage{urwchancal}
\usepackage{xcolor}
\sidecaptionvpos{figure}{t}
\usepackage{amsmath}
\usepackage{graphicx}
\usepackage{esint}
\usepackage{epstopdf}%This line makes .eps figures into .pdf - please comment out if not required.
\usepackage{rotating}
\graphicspath{{pict/}{./}}
\usepackage{bm}%
\usepackage{microtype,bm,bbm,graphicx,booktabs,times}
\newcounter{Fig}

\begin{document}

%\titlefigure{abstract}

%\title{Invariant Scattering Properties for Arbitrary Polarizations Protected by Discrete Symmetries}
%\title{Invariant scattering properties for arbitrary polarizations protected by discrete geometric symmetries}
%\title{Arbitrary-polarization scattering invariance protected by joint spatial-duality symmetries}
\title{Ideal Kerker scattering by homogeneous spheres: the role of gain or loss}
%\keywords{Scattering Activity; Symmetry; Nonreciprocity; Parity Conservation.}%
\author{Qingdong Yang}
%\email{Authors contributed equally to this work.}
\affiliation{School of Optical and Electronic Information, Huazhong University of Science and Technology, Wuhan, Hubei 430074, P. R. China}
\author{Weijin Chen}
%\email{Authors contributed equally to this work.}
\affiliation{School of Optical and Electronic Information, Huazhong University of Science and Technology, Wuhan, Hubei 430074, P. R. China}
\author{Yuntian Chen}
\email{yuntian@hust.edu.cn}
\affiliation{School of Optical and Electronic Information, Huazhong University of Science and Technology, Wuhan, Hubei 430074, P. R. China}
\affiliation{Wuhan National Laboratory for Optoelectronics, Huazhong University of Science and Technology, Wuhan, Hubei 430074, P. R. China}
\author{Wei Liu}
\email{wei.liu.pku@gmail.com}
\affiliation{College for Advanced Interdisciplinary Studies, National University of Defense
Technology, Changsha, Hunan 410073, P. R. China}

\begin{abstract}

We reexamine a recent work [Phys. Rev. Lett. \textbf{125}, 073205 (2020)] that investigates how the optical gain or loss (characterized by isotropic complex refractive indexes) influences the ideal Kerker scattering of exactly zero backward scattering. There it has been rigourously proved that, for non-magnetic homogeneous spheres with incident plane waves, either gain or loss prohibits such ideal Kerker scattering, provided that only electric and magnetic multipoles of a specific order are present and contributions from  other multipoles can all be made precisely zero. Here we reveal that, when two multipoles of a fixed order are perfectly matched in terms of both phase and magnitude,  multipoles of at least the next two orders cannot possibly be tuned to be all precisely zero or even perfectly matched, and consequently cannot directly produce ideal Kerker scattering. Moreover, we further demonstrate that, when multipoles of different orders are simultaneously taken into consideration, the loss or gain can serve as a helpful rather than harmful contributing factor, for the eliminations of backward scattering.
\end{abstract}

\maketitle

\section{Introduction}
\label{section1}
The original Kerker scattering of zero backward scattering was firstly proposed for homogeneous magnetic spheres with equal electric permittivity and magnetic permeability $\epsilon=\mu$~\cite{Kerker1983_JOSA}. Such a proposal had not attracted much attention for a long time, mainly due to the scarcity of magnetic materials, especially at the high frequency spectral regimes. In the past decade, thanks to the explosive developments of metamaterials and metasurfaces, the underlying core concept of optically-induced magnetism in non-magnetic structures has invigorated and completely transformed Kerker's original proposal (see reviews~\cite{jahani_alldielectric_2016,KUZNETSOV_Science_optically_2016,LIU_2018_Opt.Express_Generalized} and references therein). The fusion of optically-induced magnetism with Kerker scattering has rendered new perspectives for photonic studies concerning not only scattering of individual particle or their finite clusters, but also extended periodic or aperiodic structures ~\cite{jahani_alldielectric_2016,KUZNETSOV_Science_optically_2016,LIU_2018_Opt.Express_Generalized,SMIRNOVA_Optica_multipolar_2016,CHEN_Rep.Prog.Phys._review_2016,STAUDE_NatPhoton_metamaterialinspired_2017,KRUK2017ACSPhotonics,YANG2017PhysicsReports,DING_Rep.Prog.Phys._gradient_2017}.
Moreover, this significantly broadened concept of Kerker scattering has rapidly penetrated into other disciplines of photonics, revealing hidden connections between seemingly unrelated concepts and demonstrations~\cite{CHEN_2019__Singularities,CHEN_ACSOmega_Global,POSHAKINSKIY_2019_Phys.Rev.X_Optomechanical,ALAEE_Phys.Rev.Lett._Quantum,ZHU_Phys.Rev.Lett._Extraordinary,FERNANDEZ-CORBATON_2013_Opt.ExpressOE_Forwarda,YANG_2020_ACSPhotonics_Electromagnetic}.

In the original proposal for homogenous spheres with $\epsilon=\mu$, electric and magnetic multipoles of all orders are automatically perfectly matched in terms of both phase and magnitude~\cite{LEE_Phys.Rev.A_Reexamination}, leading to ideal Kerker scattering of exactly zero backward scattering~\cite{Kerker1983_JOSA}. Nevertheless, for demonstrations relying on optically-induced magnetism with non-magnetic structures ($\mu=1$), it is rather challenging, if not impossible, to precisely match all multipoles simultaneously,  ending up with only significantly suppressed but not exactly zero backward scattering~\cite{jahani_alldielectric_2016,KUZNETSOV_Science_optically_2016,LIU_2018_Opt.Express_Generalized}. Quite recently, Olmos-Trigo \textit{et al.} revisits the simplest case of a non-magnetic isotropic and homogeneous sphere with incident plane waves, and concludes that: ideal zero backward scattering is achievable only for materials without gain or loss (characterized by real refractive indexes)~\cite{OLMOS-TRIGO_Phys.Rev.Lett._Kerker}; extra gain or loss inhibits such ideal Kerker scattering.  Besides the proved feasible perfect matching of electric and magnetic multipoles of one specific fixed order, the validity of the conclusion resides on an extra assumption that magnitudes of multipoles of all other order can be simultaneously tuned to be perfectly zero.  For general discussions of optical properties such as scattering and absorption cross sections, it is physically legitimate to take into consideration only those dominant contributing multipole terms and drop other minor ones (such as the widely adopted dipole approximation).  While for the investigation into the extreme case of ideal zero backward scattering, those minor multipole terms cannot be simply discarded unless they are exactly zero or also perfectly matched in a similar fashion.

In this work we show that, despite the previously proved fact that multipoles of a fixed order can be perfectly matched in the absence of loss or gain~\cite{OLMOS-TRIGO_Phys.Rev.Lett._Kerker}, the contributions from multipoles of at least the next two orders cannot be simultaneously tuned to be all zero or perfectly matched. In other words, ideal Kerker scattering of exact zero backward scattering is not directly achievable through matching a pair of multipoles of one specific order only. We further reveal that when multipoles of different orders are all taken into consideration, loss or gain should be employed rather than avoided for the eliminations of backward scattering. It is shown that, at the presence of multipoles of various orders, the absence of backward scattering can be obtained through tuning the refractive index on the complex plane, breaking the connection between zero backscattering and helicity conservation.

\section{Formulas and analysis for ideal Kerker scattering of exactly zero backward scattering}
\label{section2}

For the scattering of incident linearly polarized plane waves (wavelength $\lambda$ and angular wave-number $k=2\pi/\lambda$) by homogeneous non-magnetic spheres (isotropic refractive index $m$, radius $R$ and normalized geometric parameter $x=kR$), the scattered fields can be expanded into a series of electric and magnetic multipoles of order $l$ ($l=1$ corresponds to dipoles). They are characterized respectively by complex Mie coefficients $a_l$ and $b_l$~\cite{hulst_light_1957,Bohren1983_book}:
%-----------------------------
\begin{equation}
a_{l}=\frac{1}{2}\left(1-e^{-2 i \alpha_l}\right),  ~b_{l}=\frac{1}{2}\left(1-e^{-2 i \beta_l}\right),
\end{equation}
%------------------------------
where $\alpha_l$ and $\beta_l$ are complex phase angles (they are real when $m$ is real).  Those phase angles can be obtained through the following relations~\cite{hulst_light_1957}:
%-----------------------------------------
\begin{equation}\begin{aligned}
\tan \alpha_{l} &=-\frac{{S}_{l}^{\prime}({mx}) {S}_{l}({x})-{mS}_{l}({mx}) {S}_{l}^{\prime}({x})}{{S}_{l}^{\prime}({mx}) {C}_{l}({x})-{mS}_{l}({mx}) {C}_{l}^{\prime}({x})}, \\
\tan \beta_{l} &=-\frac{{m} {S}_{l}^{\prime}({mx}) {S}_{l}({x})-{S}_{l}({mx}) {S}_{l}^{\prime}({x})}{{mS}_{l}^{\prime}({mx}) {C}_{l}({x})-{S}_{l}({mx}) {C}_{l}^{\prime}({x})}.
\end{aligned}\end{equation}
%--------------------------------------
Here the upper prime $^\prime$ denotes first-order derivative with respect to the entire argument in the bracket; $S_{{l}}(z)=zj_{{l}}(z)$ and $C_{{l}}(z)=-zy_{{l}}(z)$ are Riccati-Bessel functions; $j_{{l}}(z)$ and $y_{{l}}(z)$ are spherical Bessel functions of the first and second kinds.

With $a_l$ and $b_l$ obtained, the total scattering efficiency can be calculated through~\cite{hulst_light_1957,Bohren1983_book}:
%=========================
\begin{equation}
Q_{\mathrm{sca}}=\frac{2}{x^{2}} \sum_{l=1}^{\infty}(2 l+1)\left(\left|a_{l}\right|^{2}+\left|b_{l}\right|^{2}\right)
\end{equation}
%==============================
and the ideal Kerker scattering in terms of backward scattering efficiency $Q_{\rm{b}}$ can be expressed as~\cite{hulst_light_1957,Bohren1983_book}:
%----------------------------
\begin{equation}
\label{back}
Q_{\rm{b}}=\frac{1}{x^{2}}|\sum_{l=1}^{\infty}(2l+1)(-1)^{l}\left(a_{l}-b_{l}\right)|^{2}=0.
\end{equation}
%----------------------------
Equation~(\ref{back}) has infinite sets of possible solutions, and what is discussed in Ref.~\cite{OLMOS-TRIGO_Phys.Rev.Lett._Kerker} is actually the following very special scenario:
%%---------------------------
\begin{subequations}%\label{solution1}
\begin{align}
 & a_{l}=b_{l}\neq 0, \quad l=l_{0}; \label{solutiona} \\
 & a_{l}=b_{l}=0, \quad l \neq l_{0}, \label{solutionb}
\end{align}
\end{subequations}
%--------------------------------
where $l_{0}$ is an arbitrary natural number and a pair of multipoles of order $l_{0}$ are perfectly matched as shown in Eq.~(\ref{solutiona}). The significant contribution from Ref.~\cite{OLMOS-TRIGO_Phys.Rev.Lett._Kerker} is to prove rigourously that Eq.~(\ref{solutiona}) has a solution only when $m$ is real, meaning that at the presence of loss or gain multipoles of the same order cannot be ideally matched. Despite this seminal contribution, there is a problem that in Ref.~\cite{OLMOS-TRIGO_Phys.Rev.Lett._Kerker} it has not been discussed if Eqs.~(\ref{solutiona}) and~(\ref{solutionb}) are really compatible.  Such   discussions concerning compatibility are vitally important, since Eq.~(\ref{solutiona}) alone dose not necessarily lead to ideal Kerker scattering of precise zero backscattering.

\section{Mismatch among multipoles of three successive orders}
\label{section2}

In this section, we aim to prove that Eqs.~(\ref{solutiona}) and~(\ref{solutionb}) are not exactly compatible, thus proving that Ideal Kerker scattering of exact zero backward scattering is actually inaccessible through matching multipoles of a specific order only. For all our following discussions, the obviously trivial scenario of $m=1$ (we assume the background media is air of index $1$ throughout our study) or $R=0$ is excluded.  For another special case of zero index $m=0$, the Mie coefficients can be simplified as (as $m\rightarrow0$)~\cite{hulst_light_1957,Bohren1983_book}:
%----------------------------------
\begin{equation}
a_l=\frac{S_l(x)}{T_l(x)}, ~b_l=\frac{S_l^{\prime}(x)S_l(mx)}{T_l^{\prime}(x)S_l(mx)},
\end{equation}
%-----------------------------------
where $T_l(x)=xh_n^{(1)}(x)$, and $h_n^{(1)}(x)$ is spherical Hankel function of the first kind.  Since $S_l(mx)\rightarrow0$ when $m\rightarrow0$, we get a definite $a_l$ but indefinite $b_l$ (L'H\^{o}pital's rule will not help to make $b_l$ definite, since the zero term in the numerator and denominator is the same~\cite{ALEKSANDROV__Mathematics}). What means by this is that for $m=0$, there are no definite scattering properties for ideally monochromatic plane waves. Physical investigations can be implemented only after considering simultaneously the dispersion of the index and the spectrum of the incident waves. Consequently, the zero index scenario is also excluded in the following analysis.
%-----------------------------------

It has been rigorously proved that the solutions of Eq.~(\ref{solutiona}) satisfy either of the following equations:
%%---------------------------
\begin{subequations}%\label{solution1}
\begin{align}
 & S_{{l_0}}(mx)=0; \label{solutionc} \\
 & S_{{l_0}}^{\prime}(mx)=0, \label{solutiond}
\end{align}
\end{subequations}
%--------------------------------
which do not have a common solution according to the Brauer-Siegel theorem~\cite{WATSON__Treatise,OLMOS-TRIGO_ArXiv200312812Phys._Unveiling}. Similarly, to prove that then multipoles of all other orders ($l\neq  l_0$) cannot all be perfectly matched (of which that other multipoles cannot be tuned to be all zero is merely a special scenario), it is more than sufficient to prove that there exists one multipole order $l_1$ ($l_1\neq l_0$) for which:
%----------------------------------
\begin{equation}
\label{critiaria}
S_{{l_1}}(mx) \cdot S_{{l_1}}^{\prime}(mx)\neq 0.
\end{equation}
%-----------------------------------
Obviously, Eq.~(\ref{critiaria}) ensures that $a_{{l_1}}\neq b_{{l_1}}$, meaning that Eq.~(\ref{solutionb}) cannot be simultaneously met.

According to the following recurrence relations of Riccati-Bessel functions~\cite{WATSON__Treatise}:
%------------------------------------------------
\begin{subequations}%\label{solution1}
\begin{align}
  S_{l_{0}+1}^{\prime}(m x)=-\frac{l_{0}+1}{m x} S_{l_{0}+1}(m x)+S_{l_{0}}(m x) ; \label{recurrence1} \\
  S_{l_{0}}^{\prime}(m x)=\frac{l_{0}+1}{m x} S_{l_{0}}(m x)-S_{l_{0}+1}(m x), \label{recurrence2}
\end{align}
\end{subequations}
%------------------------------------------------

$\blacksquare$  ~When $S_{{l_0}}(mx)=0$ and $S_{{l_0}}^{\prime}(mx)\neq0$:
according to Eq.~(\ref{recurrence2}), we obtain $S_{l_{0}+1}(m x)\neq 0$. This together with Eq.~(\ref{recurrence1}) leads to $S_{l_{0}+1}^{\prime}(m x)\neq 0$.  As a result, Eq.~(\ref{critiaria}) is satisfied at least for $l_1=l_0+1$, securing that $a_{l_0+1} \neq b_{l_0+1}$.

$\blacksquare$  ~When $S_{{l_0}}(mx)\neq 0$ and $S_{{l_0}}^{\prime}(mx)=0$: also according to Eq.~(\ref{recurrence2}), we get $S_{l_{0}+1}(m x)\neq 0$. Nevertheless, according to Eq.~(\ref{recurrence1}), $S_{l_{0}+1}^{\prime}(m x)=0$ if the following conditions can be met:
%----------------------------------
\begin{equation}
\label{critiaria2}
S_{{l_0+1}}(mx) =\pm S_{{l_0}}(mx), ~ l_{0}+1=\pm mx.
\end{equation}
%-----------------------------------
%It means that both multipoles of order $l_0$ and $l_0+1$ can be perfectly matched:  $a_{l_0,l_0+1} = b_{l_0, l_0+1}$.     
Nevertheless,  following the same logic, extending the multipole matching to the next order $l_{0}+2$ requires:
%----------------------------------
\begin{equation}
\label{critiaria3}
S_{{l_0+2}}(mx) =\pm S_{{l_0+1}}(mx), ~ l_{0}+2=\pm mx.
\end{equation}
%-----------------------------------
It is quite obvious that Eqs.~(\ref{critiaria2}) and~(\ref{critiaria3}) can not be simultaneously satisfied [$mx$ cannot be both $\pm (l_0+1)$ and $\pm (l_0+2)$], and thus multipole mismatch happens at least for $l_1=l_0+2$: $a_{l_0+2} \neq  b_{l_0+2}$.

Those arguments above, consistent with a recent study~\cite{OLMOS-TRIGO_ArXiv200312812Phys._Unveiling}, confirm that when multipoles of a specific order $l_0$ is perfectly matched in a nontrivial way [Eq.~(\ref{solutiona})], the scattering contributions from multipoles of at least the next two successive orders ($l_0+1$ and $l_0+2$) cannot be simultaneously  tuned to be zero or matched. In other words, prefect matching of multipoles of one specific order does not guarantee ideal zero backward scattering.

%-------------------------------------------------------------------------------
\begin{figure}
\centerline{\includegraphics[width=8.9cm]{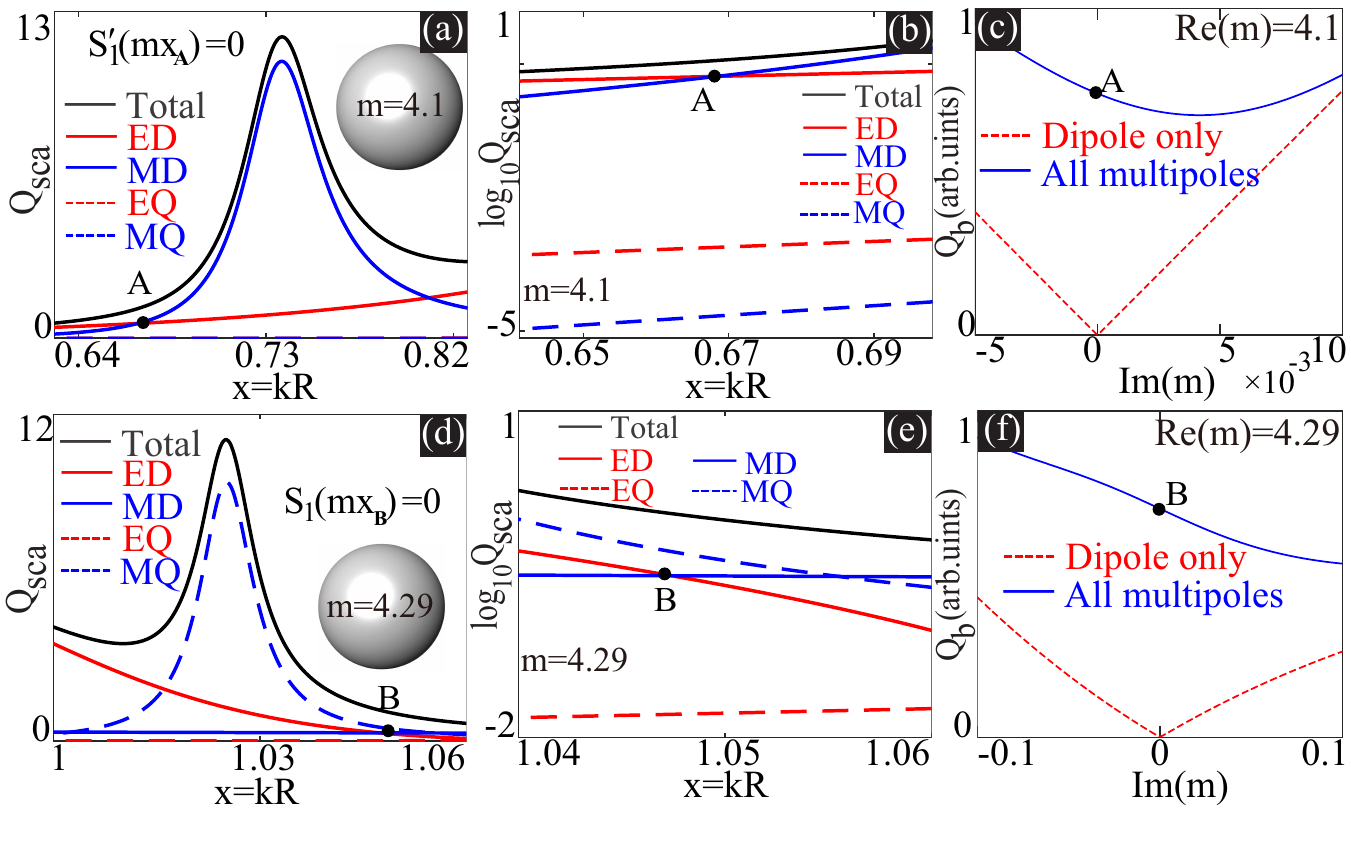}}\caption{\small Scattering spectra (both total scattering and those contributed by different multipoles are included) are shown in (a) and (b) for $m=4.1$, and in (d) and (e) for $m=4.29$. Here (b) and (e) are small parts of (a) and (b) respectively, which are close to the dipole matching points and enlarged for clarity. For each scenario, ther is a indicated point ($x_\mathbf{A}\approx 0.6684$ and $x_\mathbf{B}\approx 1.0472$) where the dipoles are perfectly matached. (c) and (f) Dependence of $Q_{\rm{b}}$ on Im($m$), at $x_\mathbf{A}$ with fiexed Re($m$)$=4.1$ and at $x_\mathbf{B}$ with fiexed Re($m$)$=4.29$, respectively. In (c) and (f) two sets of results are shown, considering only dipoles or all multipoles, respectively.}
\label{fig1}
\end{figure}
%-------------------------------------------------------------------------------

\section{Effects of gain or loss on ideal Kerker scattering: non-resonant regimes}
\label{section3}

We show in Fig.~\ref{fig1} two scenarios where the electric and magnetic dipoles (ED and MD) are perfectly matched in non-resonant spectra regimes. The scattering efficiency spectra (scattering efficiency $Q_{\rm{sca}}$ versus $x=kR$) for a homogeneous sphere ($m=4.1$) are shown in Fig.~\ref{fig1}(a), where both total scattering and those contributed by different multipoles (dipoles and electric and magnetic quadrupoles: EQ and MQ) are included.  This is actually the case studied in detail in Ref.~\cite{OLMOS-TRIGO_Phys.Rev.Lett._Kerker}. The ED and MD are perfectly matched at $x_\mathbf{A}=0.6684$, where $S_{{1}}^{\prime}(mx_\mathbf{A})=0$. As argued in the last section, at $x_\mathbf{A}$ scattering from multipoles of higher orders is not exactly zero [see Fig.~\ref{fig1}(b) that shows an enlarged part of the spectra close to $x_\mathbf{A}$ in logarithmic scale], though they are much smaller than those of dipoles. For explorations of general properties like scattering and absorption cross sections, it is fine to drop those quadrupole terms but keep the dipole terms only. Nevertheless, for study of the extreme case of ideal Kerker scattering, simply discarding those higher order terms cannot be justified and could even lead to inaccurate conclusions.

To verify the claim above, we show in Fig.~\ref{fig1}(c) the dependence of the backward scattering efficiency $Q_{\rm{b}}$ at $x_\mathbf{A}$ on the imaginary part of refractive index Im($m$): real part of $m$ is fixed at Re($m$)$=4.1$;  Im($m$)$>0$ and Im($m$)$<0$ corresponds to loss and gain respectively. Here two sets of spectra are demonstrated, for which either only dipoles or multipoles of all orders are taken into consideration. It is clear from Fig.~\ref{fig1}(c) that, when only dipoles are considered, ideal Kerker scattering is achieved when $m$ is real, and any extra loss or gain would inhibit such scattering, as is the major conclusion of Ref.~\cite{OLMOS-TRIGO_Phys.Rev.Lett._Kerker}).  In sharp contrast, when all multipoles are considered, ideal Kerker scattering is not accessible at the prefect matching point of dipoles anymore.  Moreover, as shown in Fig.~\ref{fig1}(c), extra loss can be employed to further suppress the backward scattering, serving as a friend rather than a foe for the Kerker scattering.
Another scenario of perfect dipole matching at $x_\mathbf{B}=1.0472$ for $m=4.29$ is summarized in Figs.~\ref{fig1}(d)-(f), for which the other perfect matching condition is satisfied: $S_{{1}}(mx_\mathbf{A})=0$. Here the effects of higher order multipoles are even more pronounced [see Fig.~\ref{fig1}(f)], since the magnitudes of dipoles and higher multipoles are comparable [see Fig.~\ref{fig1}(e)].

%-------------------------------------------------------------------------------
\begin{figure}
\centerline{\includegraphics[width=8.9cm]{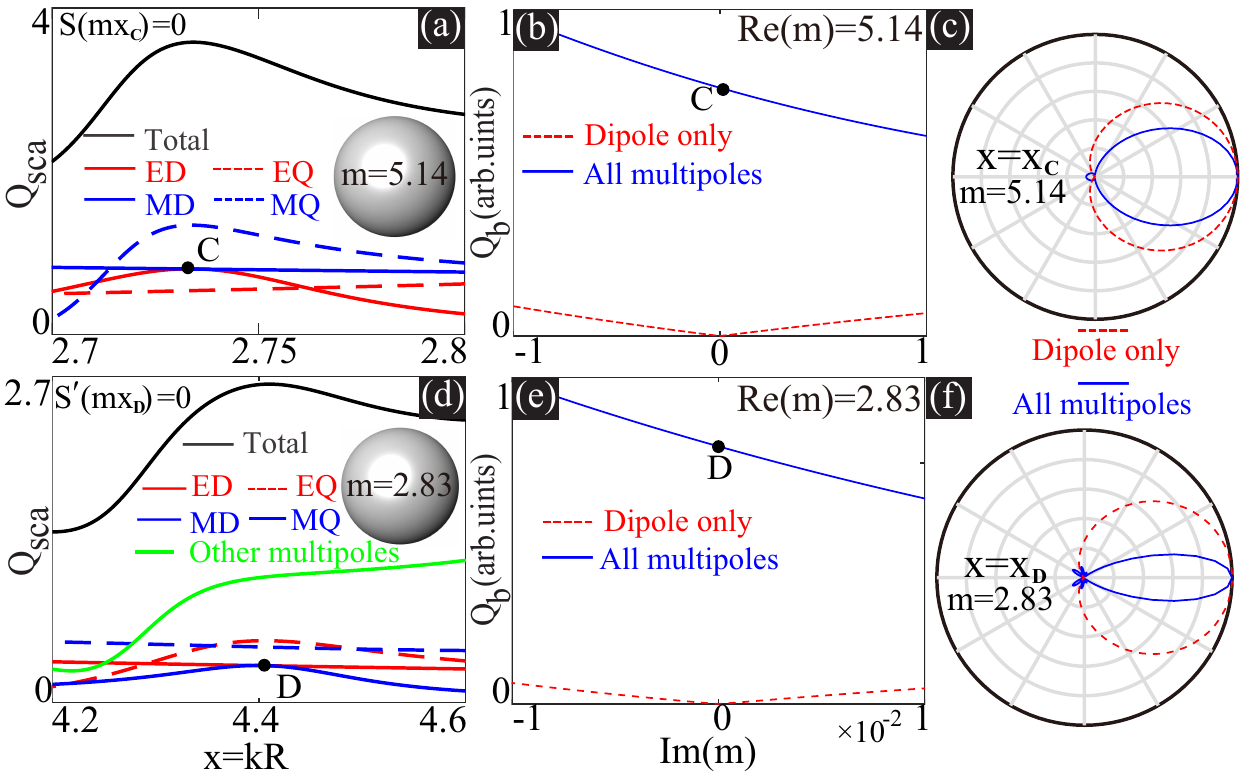}}\caption{\small Scattering spectra are shown in (a) for $m=5.14$, and in (d) for $m=2.83$. For each scenario, there is a indicated point ($x_\mathbf{{C}}\approx 2.7366$ and $x_\mathbf{D}\approx 4.4123$) where the dipoles are perfectly matached. (b) and (e) Dependence of $Q_{\rm{b}}$ on Im($m$), at $x_\mathbf{C}$ with fixed Re($m$)$=5.14$ and at $x_\mathbf{D}$ with fiexed Re($m$)$=2.83$, respectively. (c) and (f) The 2D angular scattering patterns (on the plane parrael to both polarization and incidnet directions) at $x_\mathbf{C}$ with $m=5.14$ and at $x_\mathbf{D}$ with $m=2.83$, respectively. In (b)-(c) and (e)-（f) two sets of results are shown, considering only dipoles or all multipoles, respectively.}
\label{fig2}
\end{figure}
%-------------------------------------------------------------------------------

\section{Effects of gain or loss on ideal Kerker scattering: resonant regimes}
\label{section4}

In the last section, we discuss only the perfect dipole matching at the non-resonant regimes, where not only the backward scattering is suppressed, but also the overall scattering is small. Such scattering is of very limited significance, since what is widely required in photonics is suppressed backward scattering accompanied by large total scattering~\cite{jahani_alldielectric_2016,KUZNETSOV_Science_optically_2016,LIU_2018_Opt.Express_Generalized}. In this section, we move to the resonant regimes where the dipoles can be perfectly matched.  Two such scenarios are summarized in  Fig.~\ref{fig2}, where the conditions of $S_{{1}}(mx_\mathbf{C})=0$ and $S_{{1}}^{\prime}(mx_\mathbf{D})=0$ are satisfied, in Figs.~\ref{fig2}(a)-(c) ($x_\mathbf{C}\approx 2.7366$, $m=5.14$) and Figs.~\ref{fig2}(d)-(f) ($x_\mathbf{D}\approx 4.4123$, and $m=2.83$), respectively.  In Fig.~\ref{fig2}, besides the scattering spectra [Figs.~\ref{fig2}(a) and (d)] and dependence of $Q_{\rm{b}}$ on Im($m$) [Figs.~\ref{fig2}(b) and (e)], we show also the two dimensional (2D) scattering patterns (on the plane parrel to both the incident and polarization directions of the independent plane waves) at the dipole matching points [Figs.~\ref{fig2}(c) and (f)]. As indicated by the scattering spectra, the scatterings by the higher order multipoles are rather strong, which completely ruins the ideal Kerker scattering [see Figs.~\ref{fig2}(b) and (e)]. Similar to what is shown in Fig.~\ref{fig1}, when all multipoles are considered, extra loss can be employed to further suppress the backward scattering, serving as a constructive rather than a destructive factor for demonstrations of Kerker scattering.

%-------------------------------------------------------------------------------
\begin{figure}
\centerline{\includegraphics[width=8.9cm]{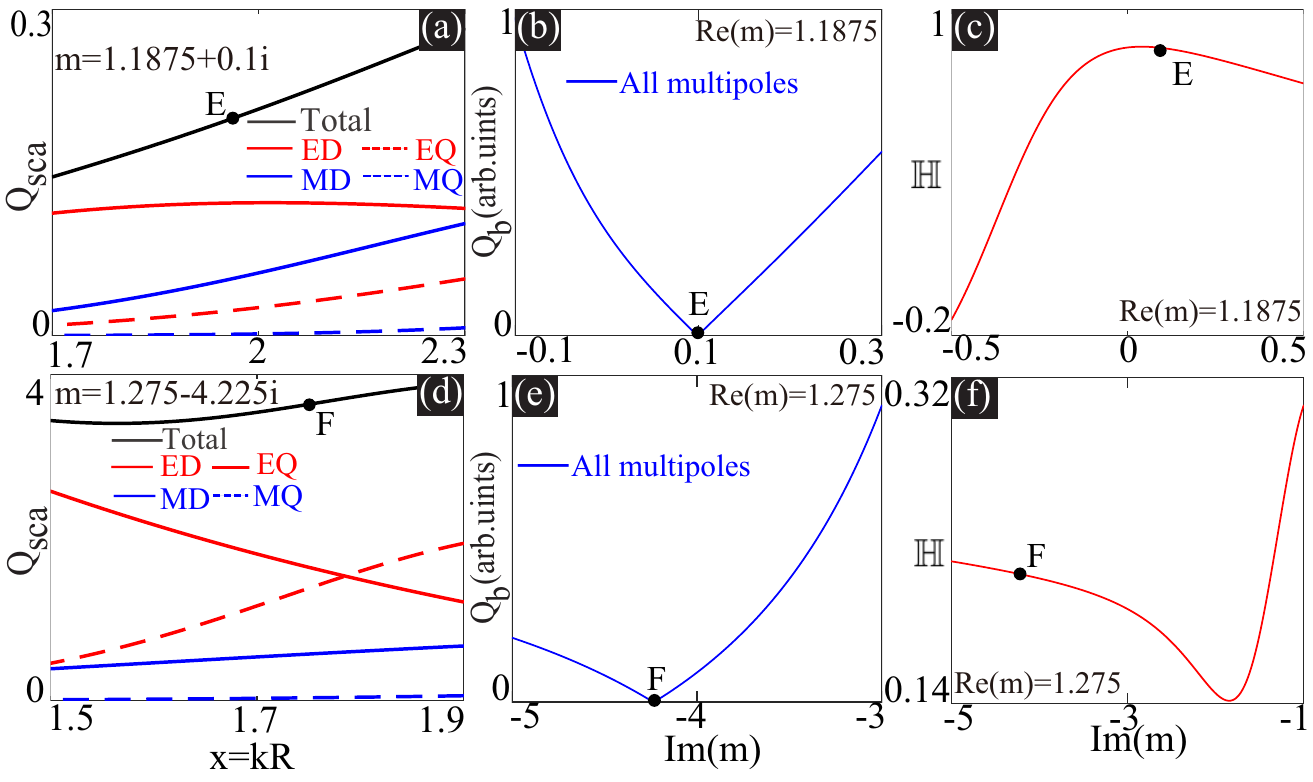}}\caption{\small Scattering spectra are shown in (a) for $m=1.1875+0.1i$ (with loss), and in (d) for $m=1.275-4.225i$ (with gain). For each scenario, ther is a indicated point ($x_\mathbf{E}\approx 1.9591$ and $x_\mathbf{F}\approx 1.7492$) where the backward scattering is eliminated, as confirmed in (b) and (e).  Dependence of $Q_{\rm{b}}$ [(b) \& (e)] and $\mathbb{H}$ [(c) \& (f)] on Im($m$),
at $x_\mathbf{E}$ with fixed Re($m$)$=1.1875$ and at $x_\mathbf{F}$ with fixed Re($m$)$=1.275$, respectively.}
\label{fig3}
\end{figure}
%-------------------------------------------------------------------------------

\section{Kerker scattering without multipole matching of any specific order}
\label{section4}

We have confirmed in the last sections, by both mathematical analysis and numerical calculations, that perfect matching of multipoles of a specific order does not necessarily produce ideal Kerker scattering due to noneligible higher order multipoles. Moreover, those higher order terms would make the extra gain or loss a constructive factor for further suppressions of the backward scattering.  Now we come back to Eq.~(\ref{back}), the solution of which does not really require multipole matching of any specific order [such as those shown in Eqs.~(\ref{solutiona}) and (\ref{solutionb})], but can be obtained through fully destructive interferences among multipoles involving several orders along the backward direction. To confirm this point, we show two such scenarios with loss or gain in Fig.~\ref{fig3}, where the Kerker scattering is observed in Figs.~\ref{fig3}(a) and (b) ($x_\mathbf{E}\approx 1.9591$, $m=1.1875+0.1i$), and in Figs.~\ref{fig3}(c) and (d) ($x_\mathbf{F}\approx 1.7492$ for $m=1.275-4.225i$), respectively.   Figures~\ref{fig3}(a) and (b) show that there is no non-trivial perfect multipole matching ($a_{l}=b_{l}\neq 0$) at the indicated positions, despite which the Kerker scattering can still be achieved [see Figs.~\ref{fig3}(b) and (e) at $x_\mathbf{E}$ and $x_\mathbf{F}$, respectively].  Moreover, the dependence of $Q_{\rm{b}}$ on Im($m$) [Figs.~\ref{fig3}(b) and (e)] can confirm that the selected loss or gain is vitally important for such achievement, as a little detuning from them would immediately ruin the Kerker scattering.  For both scenarios, it is quite obvious that to fix the index to be real actually harmful for the suppressions of backward scatterings.

It has been rigourously proved that $n$-fold ($n\geq3$) rotation symmetry together with helicity conservation would automatically guarantee ideal Kerker scattering of zero backward scattering~\cite{FERNANDEZ-CORBATON_2013_Opt.ExpressOE_Forwarda,YANG_2020_ArXiv200613466Phys._Symmetry}. For homogenous sphere scattering with incident plane waves, the rotation symmetry is secured ($n=\infty$) and the helicity conservation requires the multipole matching of all orders. Consequently, Kerker scattering obtained through perfect matching of multipoles at each order are inextricably connected through helicity conservation, as is confirmed in Ref.~\cite{OLMOS-TRIGO_Phys.Rev.Lett._Kerker}. Nevertheless,  we have shown  in the last section that Kerker scattering is also achievable without multipole matching of any specific order, for which it is expected that the connection between Kerker scattering and helicity conservation would be broken. To confirm this, we further show the dependence of helicity conservation factor $\mathbb{H}$ on Im($m$) in Figs.~\ref{fig3}(c) and (f).  Here $\mathbb{H}$ is defined as~\cite{OLMOS-TRIGO_Phys.Rev.Lett._Kerker,OLMOS-TRIGO_Opt.Lett.OL_Asymmetry}:
%=====================
\begin{equation}
\mathbb{H}=\frac{\sum_{l=1}^{\infty}(2 l+1) \rm{Re}\left\{a_{\textit{l}} b_{\textit{l}}^{*}\right\}}{\sum_{l=1}^{\infty}(l+1/2)\left(\left|a_{l}\right|^{2}+\left|b_{l}\right|^{2}\right)}.
\end{equation}
%=====================
Here $\mathbb{H}=1$ corresponds to ideal helicity conservation, which means that for an incident circularly-polarized plane waves, the waves scattered along all directions are also circularly-polarized of the same handedness (including the special case of zero scattering)~\cite{FERNANDEZ-CORBATON_2013_Opt.ExpressOE_Forwarda,CHEN_ACSOmega_Global,FERNANDEZ-CORBATON_2013_Phys.Rev.Lett._Electromagnetica,YANG_ArXiv200701535Phys._Scatteringa}. A comparison between Figs.~\ref{fig3}(c) and (f) and Figs.~\ref{fig3}(b) and (e) can confirm that there is no connection between the Kerker scattering and helicity conservation, since $\mathbb{H}$ is far from unity at the indicated Kerker scattering points ($\mathbb{H}_{\mathbf{E}}=0.857$, $\mathbb{H}_{\mathbf{F}}=0.2115$). In other words, rotation symmetry and helicity conservation lead to zero backward scattering, while rotation symmetry and zero backward scattering does not necessarily imply helicity conservation.

\section{Conclusions and Discussions}

To conclude, we prove that perfectly matching electric and magnetic multipoles of a specific order does not necessarily produce ideal Kerker scattering of exact zero backward scattering, since no matter how small the contributions from other multipoles are, they can never be made to be all zero or perfectly matched. In other words, to obtain zero backward scattering, we cannot just consider multipoles of a specific order, but all contributing ones that are not exactly zero. It is further demonstrated that when multipoles of various order are simultaneously considered, loss or gain can be employed for suppressions of backward scattering, serving as beneficial rather than detrimental elements for the realizations of ideal Kerker scattering. When the Kerker scattering is achieved through the destructive interference among multipoles of several orders in the backward direction, rather than perfect multipole matching of each order, it is not synonymous with helicity conservation any more.

There are several significant points worth emphasizing at the end: \\

(i) For numerical demonstrations of perfect multipole matching, we discuss only dipoles while the principles revealed are applicable for multipoles of any order. \\

(ii) In this study, we only discuss Kerker scattering of zero backward scattering (first Kerker scattering).  For the second Kerker scattering of zero forward scattering, despite the inevitable involvement of gain materials as required by optical theorem, multipoles of various order rather than a specific order should be taken into considerations simultaneously, as has been implemented in this work. \\

(iii) Is ideal Kerker scattering of exact zero backward scattering achievable, in a rigorously mathematical sense, with homogenous non-magnetic spheres? The answer is we do not know. It is well known that for arbitrary algebraic equations of order $L$ ($\sum_{l=0}^{L}c_lx^{L-l}=0$, for which $L$ is a finite natural number and $c_l$ are complex constant coefficients), the fundamental theorem of algebra secures that there is at least one solution on the complex $x$-plane~\cite{ALEKSANDROV__Mathematics}. Nevertheless, Eq.~(\ref{back}) is a transcendental rather than algebraic equation, of which the existence of exact solution on the complex plane is not definite. Such a transcendental equation can be only tackled through numerical analysis and thus numerical errors make it impossible to decide if the Kerker scattering demonstrated in  Fig.~\ref{fig3} is ideal or not in a mathematical sense. \\

(iv) If the exact solution of Eq.~(\ref{back}) exists, the chances of this solution being complex are much higher than it being purely real (real axes cover a tiny part of the complex plane). If exact solution does not exist, the backward scattering is minimized more probably at complex arguments rather than at purely real ones. As a result, gain or loss are definitely helpful rather than harmful for the realizations of ideal Kerker scattering or backward scattering suppressions.  \\

(v) Discussing the exact solution of Eq.~(\ref{back}) (and thus ideal Kerker scattering) is interesting and meaningful only mathematically. From a physical perspective, such an exploration is of very little significance, if not of no significance at all.  This is because for realistic observations, there is no absolute boundary between exactly zero and approximately zero, which is highly dependent on the resolutions of different equipments. Moreover, when the scattering intensity gets smaller and smaller,  the optical regime we study will shift from wave optics to quantum optics, where the quantum fluctuations would play a noneligible role~\cite{BERRY_Int.Conf.Singul.Opt._Much}.  Then wave optics and thus Eq.~(\ref{back}) itself breaks down and thus it becomes meaningless to discuss its exact solution.\\

\section*{acknowledgement}
We acknowledge the financial support from National Natural Science
Foundation of China (Grant No. 11874026, and 11874426), and the Outstanding Young Researcher Scheme of National University of Defense Technology. W. L. thanks
Dr. Jorge Olmos-Trigo for his comments on this work.

%Q. Yang and W. Chen contributed equally to this work.

%load xkeyval.sty
%\bibliographystyle{osajnl}
%\bibliography{References_scattering3} %the RSC's .bst file

\end{document}